\documentclass[aps,prd,twocolumn,showpacs,superscriptaddress,groupedaddress]{revtex4}  
\usepackage{graphicx}  
\usepackage{dcolumn}   
\usepackage{bm}        
\usepackage{amssymb}   

\begin{document}

\hspace{5.2in} \mbox{Fermilab-Pub-15-496-E}

\title{Measurement of the forward-backward asymmetry of $\Lambda$ and $\bar{\Lambda}$
production in $p \bar{p}$ collisions}
\affiliation{LAFEX, Centro Brasileiro de Pesquisas F\'{i}sicas, Rio de Janeiro, Brazil}
\affiliation{Universidade do Estado do Rio de Janeiro, Rio de Janeiro, Brazil}
\affiliation{Universidade Federal do ABC, Santo Andr\'e, Brazil}
\affiliation{University of Science and Technology of China, Hefei, People's Republic of China}
\affiliation{Universidad de los Andes, Bogot\'a, Colombia}
\affiliation{Charles University, Faculty of Mathematics and Physics, Center for Particle Physics, Prague, Czech Republic}
\affiliation{Czech Technical University in Prague, Prague, Czech Republic}
\affiliation{Institute of Physics, Academy of Sciences of the Czech Republic, Prague, Czech Republic}
\affiliation{Universidad San Francisco de Quito, Quito, Ecuador}
\affiliation{LPC, Universit\'e Blaise Pascal, CNRS/IN2P3, Clermont, France}
\affiliation{LPSC, Universit\'e Joseph Fourier Grenoble 1, CNRS/IN2P3, Institut National Polytechnique de Grenoble, Grenoble, France}
\affiliation{CPPM, Aix-Marseille Universit\'e, CNRS/IN2P3, Marseille, France}
\affiliation{LAL, Universit\'e Paris-Sud, CNRS/IN2P3, Orsay, France}
\affiliation{LPNHE, Universit\'es Paris VI and VII, CNRS/IN2P3, Paris, France}
\affiliation{CEA, Irfu, SPP, Saclay, France}
\affiliation{IPHC, Universit\'e de Strasbourg, CNRS/IN2P3, Strasbourg, France}
\affiliation{IPNL, Universit\'e Lyon 1, CNRS/IN2P3, Villeurbanne, France and Universit\'e de Lyon, Lyon, France}
\affiliation{III. Physikalisches Institut A, RWTH Aachen University, Aachen, Germany}
\affiliation{Physikalisches Institut, Universit\"at Freiburg, Freiburg, Germany}
\affiliation{II. Physikalisches Institut, Georg-August-Universit\"at G\"ottingen, G\"ottingen, Germany}
\affiliation{Institut f\"ur Physik, Universit\"at Mainz, Mainz, Germany}
\affiliation{Ludwig-Maximilians-Universit\"at M\"unchen, M\"unchen, Germany}
\affiliation{Panjab University, Chandigarh, India}
\affiliation{Delhi University, Delhi, India}
\affiliation{Tata Institute of Fundamental Research, Mumbai, India}
\affiliation{University College Dublin, Dublin, Ireland}
\affiliation{Korea Detector Laboratory, Korea University, Seoul, Korea}
\affiliation{CINVESTAV, Mexico City, Mexico}
\affiliation{Nikhef, Science Park, Amsterdam, the Netherlands}
\affiliation{Radboud University Nijmegen, Nijmegen, the Netherlands}
\affiliation{Joint Institute for Nuclear Research, Dubna, Russia}
\affiliation{Institute for Theoretical and Experimental Physics, Moscow, Russia}
\affiliation{Moscow State University, Moscow, Russia}
\affiliation{Institute for High Energy Physics, Protvino, Russia}
\affiliation{Petersburg Nuclear Physics Institute, St. Petersburg, Russia}
\affiliation{Instituci\'{o} Catalana de Recerca i Estudis Avan\c{c}ats (ICREA) and Institut de F\'{i}sica d'Altes Energies (IFAE), Barcelona, Spain}
\affiliation{Uppsala University, Uppsala, Sweden}
\affiliation{Taras Shevchenko National University of Kyiv, Kiev, Ukraine}
\affiliation{Lancaster University, Lancaster LA1 4YB, United Kingdom}
\affiliation{Imperial College London, London SW7 2AZ, United Kingdom}
\affiliation{The University of Manchester, Manchester M13 9PL, United Kingdom}
\affiliation{University of Arizona, Tucson, Arizona 85721, USA}
\affiliation{University of California Riverside, Riverside, California 92521, USA}
\affiliation{Florida State University, Tallahassee, Florida 32306, USA}
\affiliation{Fermi National Accelerator Laboratory, Batavia, Illinois 60510, USA}
\affiliation{University of Illinois at Chicago, Chicago, Illinois 60607, USA}
\affiliation{Northern Illinois University, DeKalb, Illinois 60115, USA}
\affiliation{Northwestern University, Evanston, Illinois 60208, USA}
\affiliation{Indiana University, Bloomington, Indiana 47405, USA}
\affiliation{Purdue University Calumet, Hammond, Indiana 46323, USA}
\affiliation{University of Notre Dame, Notre Dame, Indiana 46556, USA}
\affiliation{Iowa State University, Ames, Iowa 50011, USA}
\affiliation{University of Kansas, Lawrence, Kansas 66045, USA}
\affiliation{Louisiana Tech University, Ruston, Louisiana 71272, USA}
\affiliation{Northeastern University, Boston, Massachusetts 02115, USA}
\affiliation{University of Michigan, Ann Arbor, Michigan 48109, USA}
\affiliation{Michigan State University, East Lansing, Michigan 48824, USA}
\affiliation{University of Mississippi, University, Mississippi 38677, USA}
\affiliation{University of Nebraska, Lincoln, Nebraska 68588, USA}
\affiliation{Rutgers University, Piscataway, New Jersey 08855, USA}
\affiliation{Princeton University, Princeton, New Jersey 08544, USA}
\affiliation{State University of New York, Buffalo, New York 14260, USA}
\affiliation{University of Rochester, Rochester, New York 14627, USA}
\affiliation{State University of New York, Stony Brook, New York 11794, USA}
\affiliation{Brookhaven National Laboratory, Upton, New York 11973, USA}
\affiliation{Langston University, Langston, Oklahoma 73050, USA}
\affiliation{University of Oklahoma, Norman, Oklahoma 73019, USA}
\affiliation{Oklahoma State University, Stillwater, Oklahoma 74078, USA}
\affiliation{Oregon State University, Corvallis, Oregon 97331, USA}
\affiliation{Brown University, Providence, Rhode Island 02912, USA}
\affiliation{University of Texas, Arlington, Texas 76019, USA}
\affiliation{Southern Methodist University, Dallas, Texas 75275, USA}
\affiliation{Rice University, Houston, Texas 77005, USA}
\affiliation{University of Virginia, Charlottesville, Virginia 22904, USA}
\affiliation{University of Washington, Seattle, Washington 98195, USA}
\author{V.M.~Abazov} \affiliation{Joint Institute for Nuclear Research, Dubna, Russia}
\author{B.~Abbott} \affiliation{University of Oklahoma, Norman, Oklahoma 73019, USA}
\author{B.S.~Acharya} \affiliation{Tata Institute of Fundamental Research, Mumbai, India}
\author{M.~Adams} \affiliation{University of Illinois at Chicago, Chicago, Illinois 60607, USA}
\author{T.~Adams} \affiliation{Florida State University, Tallahassee, Florida 32306, USA}
\author{J.P.~Agnew} \affiliation{The University of Manchester, Manchester M13 9PL, United Kingdom}
\author{G.D.~Alexeev} \affiliation{Joint Institute for Nuclear Research, Dubna, Russia}
\author{G.~Alkhazov} \affiliation{Petersburg Nuclear Physics Institute, St. Petersburg, Russia}
\author{A.~Alton$^{a}$} \affiliation{University of Michigan, Ann Arbor, Michigan 48109, USA}
\author{A.~Askew} \affiliation{Florida State University, Tallahassee, Florida 32306, USA}
% ...
%
%
\collaboration{The D0 Collaboration\footnote{with visitors from
$^{a}$Augustana College, Sioux Falls, SD, USA,
$^{b}$The University of Liverpool, Liverpool, UK,
$^{c}$DESY, Hamburg, Germany,
$^{d}$CONACyT, Mexico City, Mexico,
$^{e}$SLAC, Menlo Park, CA, USA,
$^{f}$University College London, London, UK,
$^{g}$Centro de Investigacion en Computacion - IPN, Mexico City, Mexico,
$^{h}$Universidade Estadual Paulista, S\~ao Paulo, Brazil,
$^{i}$Karlsruher Institut f\"ur Technologie (KIT) - Steinbuch Centre for Computing (SCC),
D-76128 Karlsruhe, Germany,
$^{j}$Office of Science, U.S. Department of Energy, Washington, D.C. 20585, USA,
$^{k}$American Association for the Advancement of Science, Washington, D.C. 20005, USA,
$^{l}$Kiev Institute for Nuclear Research, Kiev, Ukraine,
$^{m}$University of Maryland, College Park, MD 20742, USA,
$^{n}$European Orgnaization for Nuclear Research (CERN), Geneva, Switzerland
and
$^{o}$Purdue University, West Lafayette, IN 47907, USA.
}} \noaffiliation
\vskip 0.25cm
\date{November 16, 2015}

\begin{abstract}
\noindent
We study $\Lambda$ and $\bar{\Lambda}$
production asymmetries in 
$p \bar{p} \rightarrow \Lambda (\bar{\Lambda}) X$,
$p \bar{p} \rightarrow J/\psi \Lambda (\bar{\Lambda}) X$, and
$p \bar{p} \rightarrow \mu^\pm \Lambda (\bar{\Lambda}) X$
events recorded by the D0 detector at the Fermilab 
Tevatron collider at $\sqrt{s} = 1.96$ TeV.
We find an excess of $\Lambda$'s ($\bar{\Lambda}$'s)
produced in the proton (antiproton) direction.
This forward-backward asymmetry is measured as a function of rapidity.
We confirm that the $\bar{\Lambda}/\Lambda$ production ratio,
measured by several experiments with various targets
and a wide range of energies, is a universal function of ``rapidity loss",
i.e., the rapidity difference of the beam proton and the lambda.
\end{abstract}

\pacs {13.85.Ni, 13.60.Rj}
\maketitle



\section{Introduction}
We study $p \bar{p}$ collisions at a total center-of-mass
energy $\sqrt{s} = 1.96$ TeV.
Among the particles produced in these collisions are
$\Lambda$'s and $\bar{\Lambda}$'s.
In this paper we
examine the question of whether the $\Lambda$ and $\bar{\Lambda}$
retain some memory of the proton and antiproton beam directions.
We consider the picture in which 
a strange quark produced directly in the hard scattering of point-like
partons, or indirectly in the subsequent showering, 
can coalesce with a diquark remnant
of the beam to produce a lambda particle, with the probability
increasing with decreasing rapidity difference between the proton and the lambda \cite{x,x2,x3,x4}.

The data were recorded in the D0 detector \cite{run2det, run2muon, layer0, layer0_2, layer0_3}
at the Fermilab Tevatron collider.
The full data set of $10.4$ fb$^{-1}$, collected from 2002 to 2011, is analyzed.
We choose a coordinate
system in which the $z$ axis is aligned with the proton
beam direction and
define the rapidity $y \equiv \frac{1}{2} \ln{[(E + p_z)/(E - p_z)]}$, where $p_z$ is the
outgoing particle momentum component in the $z$ direction,
and $E$ is its energy, both in the $p \bar{p}$ center-of-mass frame.
We measure the ``forward-backward asymmetry" $A_{FB}$,
i.e. the relative excess of $\Lambda$'s ($\bar{\Lambda}$'s)
with longitudinal momentum 
in the $p$ ($\bar{p}$) direction, as a function of $|y|$.
The measurements include $\Lambda$'s and $\bar{\Lambda}$'s
from all sources either directly produced or decay products of
heavier hadrons.

The $\Lambda$'s ($\bar{\Lambda}$'s) are defined as ``forward"
if their longitudinal momentum is in the $p$ ($\bar{p}$) direction.
The asymmetry 
$A_{FB}$ is defined as
\begin{eqnarray}
A_{FB} & \equiv & \frac{\sigma_F(\Lambda) - \sigma_B(\Lambda) + \sigma_F(\bar{\Lambda}) - \sigma_B(\bar{\Lambda})}
{\sigma_F(\Lambda) + \sigma_B(\Lambda) + \sigma_F(\bar{\Lambda}) + \sigma_B(\bar{\Lambda})}, 
\label{lin0}
\end{eqnarray}
where $\sigma_F(\Lambda)$ and $\sigma_B(\Lambda)$ [$\sigma_F(\bar{\Lambda})$ and
$\sigma_B(\bar{\Lambda})$] are the forward and backward cross sections of $\Lambda$
[$\bar{\Lambda}$] production.


\begin{table}
\caption{\label{numbers}
Number of reconstructed
$\Lambda$ plus $\bar{\Lambda}$ or $K_S$ with $p_T > 2.0$ GeV
in each data set.
}
\begin{ruledtabular}
\begin{tabular}{lr}
Data set & Number of events \\
\hline
(i) $p \bar{p} \rightarrow \Lambda (\bar{\Lambda}) X$ & $5.85 \times 10^5$ \\ 
(ii) $p \bar{p} \rightarrow J/\psi \Lambda (\bar{\Lambda}) X$ & $2.50 \times 10^5$ \\ 
(iii) $p \bar{p} \rightarrow \mu^\pm \Lambda (\bar{\Lambda}) X$ & $1.15 \times 10^7$ \\ 
\hline
 (i) $p \bar{p} \rightarrow K_S X$ & $2.33 \times 10^6$ \\
(ii) $p \bar{p} \rightarrow J/\psi K_S X$ & $6.55 \times 10^5$ \\
(iii) $p \bar{p} \rightarrow \mu^\pm K_S X$ & $5.34 \times 10^7$ \\
\end{tabular}
\end{ruledtabular}
\end{table}

\section{Detector and data}
The D0 detector is described
in Refs.\ \cite{run2det, run2muon, layer0, layer0_2, layer0_3}.
The collision region is surrounded by a 
central tracking system that
comprises a silicon micro-strip vertex detector and a central fiber
tracker, both located within a $1.9$~T superconducting solenoidal
magnet~\cite{run2det}, surrounded successively by the liquid argon-uranium
calorimeters, layer A of the muon system~\cite{run2muon}
(with drift chambers and scintillation trigger counters), the 
1.8~T magnetized iron toroids, and two similar 
muon detector layers B and C after the toroids. 
The designs are optimized for vertex finding, tracking, and muon
triggering and identification at pseudorapidities $|\eta|$ less than
2.5, 3.0, and 2.0 respectively. Pseudorapidity is defined as
$\eta = - \ln {\tan (\theta/2)}$, where $\theta$ is the polar
angle with respect to the proton beam direction.


We study three data sets: 
(i) $p \bar{p} \rightarrow \Lambda (\bar{\Lambda}) X$,
(ii) $p \bar{p} \rightarrow J/\psi \Lambda (\bar{\Lambda}) X$, and
(iii) $p \bar{p} \rightarrow \mu^\pm \Lambda (\bar{\Lambda}) X$,
and corresponding control samples with $K_S$ instead
of $\Lambda$ or $\bar{\Lambda}$.
Data set (i) is collected with a prescaled trigger on beam crossing (``zero bias events") or
with a prescaled trigger on energy deposited in forward luminosity counters
(``minimum bias events"). 
Data set (ii) is selected with a suite of single muon,
dimuon, and dedicated $J/\psi$ triggers, from which $J/\psi \rightarrow \mu^+ \mu^-$ candidates
in association with a $\Lambda$ or $\bar{\Lambda}$ are
reconstructed.
Data set (iii) is selected with a suite of single muon triggers,
and a $\mu$ and a $\Lambda$ are fully reconstructed offline.
Data set (i) is unbiased, while most events in data sets (ii)
and (iii) contain heavy quarks $b$ or $c$ \cite{dimu_asym3, dimu_asym4}.
Data set (iii) has the same muon triggers and muon selections as 
in Refs.\ \cite{dimu_asym3, dimu_asym4}.
In particular, the muons are required to have 
a momentum transverse to the beams $p_T > 4.2$ GeV or
$p_z > 5.4$ GeV in order to traverse the central or forward
iron toroid magnets.
The number of reconstructed $\Lambda$ plus $\bar{\Lambda}$ or $K_S$ 
in each data sample is summarized in Table \ref{numbers}.
There is no strong physics reason to require a $J/\psi$ or $\mu$ in an event:
data sets (ii) and (iii) are analyzed because they are
collected with muon or $J/\psi$ triggers, and therefore
are available and well understood, and data set (iii) is very large.
The overlaps of the three data sets are negligible.

\begin{figure}
\begin{center}
\scalebox{0.35}
{\includegraphics{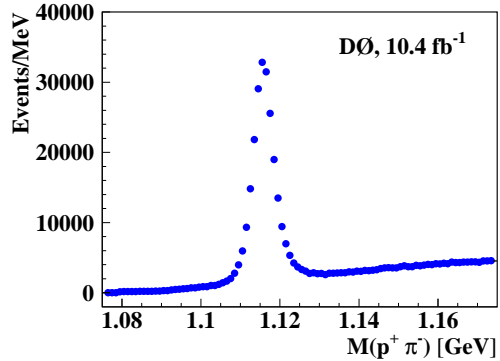}}
\caption{
Invariant mass distribution of $\Lambda \rightarrow p \pi^-$
candidates for $0.0 < y < 1.0$, 
muon charge $q = +1$, solenoid magnet polarity $-1$, 
and toroid magnet polarity $-1$, for the
$p \bar{p} \rightarrow \mu^\pm \Lambda (\bar{\Lambda}) X$ data.
Other selection requirements are given in the text.
}
\label{fit_0_1}
\end{center}
\end{figure}

\begin{figure}
\begin{center}
\scalebox{0.40}
{\includegraphics{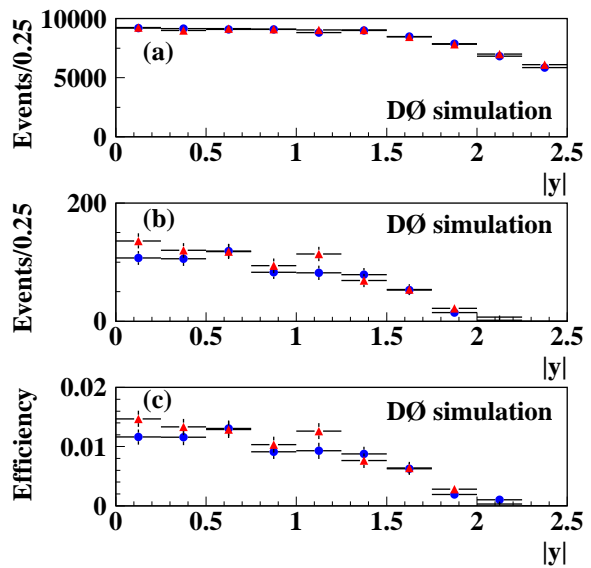}}
\caption{
(color online)
Distributions of (a) generated and (b) reconstructed         
$\Lambda$'s (blue circles) and $\bar{\Lambda}$'s (red triangles),
and (c) the corresponding efficiencies, for $p_T > 2.0$ GeV,
from QCD simulations of
inclusive $p \bar p$ collisions containing a minimum parton transverse
energy $E^{\rm min}_T > 20$~GeV. For details of the
simulation see Ref. \cite{2010}.
}
\label{eff_MC}
\end{center}
\end{figure}

The $\Lambda$'s, $\bar{\Lambda}$'s, and $K_S$'s are reconstructed
from pairs of oppositely charged tracks with a common
vertex ($V^0$). 
Each track is
required to have a non-zero impact parameter in the transverse plane (IP)
with respect to the primary $p \bar{p}$
vertex with a significance of at least two standard deviations,
and the $V^0$ projected to its point of closest approach
is required to have an IP significance less than
three standard deviations. The distance in the transverse plane
from the primary $p \bar{p}$ vertex to the $V^0$ vertex is required to be
greater than 4 mm. The $V^0$ is required to have
$2.0 < p_T < 25$ GeV and $|\eta| < 2.2$.
For $\Lambda$'s and $\bar{\Lambda}$'s, the 
proton (pion) mass is assigned to the daughter track with
larger (smaller) momentum.
This assignment is nearly always correct because the decay 
$\Lambda \rightarrow p \pi^-$ is barely above threshold.
We require that the $V^0$ daughter tracks 
not be identified as a muon.
An example of an invariant mass distribution $M(\Lambda \rightarrow p \pi^-)$
is presented in Fig.\ \ref{fit_0_1}.
The D0 detector $|y|$ acceptance is narrower than the 
lambda production rapidity plateau as shown in Fig.\ \ref{eff_MC}.

Control samples with $K_S$ are analyzed in the same manner as 
the corresponding sets with $\Lambda$ or $\bar{\Lambda}$, except that
the track with larger momentum is assigned
the pion mass instead of the proton mass. Note that we
count the decays $K_S \rightarrow \pi^+ \pi^-$ and
$K_S \rightarrow \pi^- \pi^+$ separately, where the first pion has the
larger total momentum. This way the former decay has kinematics similar
to $\Lambda$ decays, while the latter is similar to 
$\bar{\Lambda}$ decays.
The $p \bar{p}$ collisions produce $K^0$'s and $\bar{K}^0$'s that
we observe as resonances in invariant mass distributions of 
$K_S \rightarrow \pi^+ \pi^-$ decays.
Since this final state does not distinguish the parent $K^0$
from $\bar{K}^0$ (neglecting CP violation), 
$K_S$ decays do not distinguish the $p$ and $\bar{p}$
directions,
have no physics asymmetries, and
so constitute a control sample to study detector effects.

\section{Raw asymmetries and detector effects}

We observe $\Lambda$'s and $\bar{\Lambda}$'s through their decays
$\Lambda \rightarrow p \pi^-$ and $\bar{\Lambda} \rightarrow \overline{p} \pi^+$.
We obtain the numbers $N_F(\Lambda)$ and $N_B(\Lambda)$ 
[$N_F(\bar{\Lambda})$ and $N_B(\bar{\Lambda})$]
of reconstructed $\Lambda$'s [$\bar{\Lambda}$'s] in the ``forward" and ``backward"
categories, respectively, in each bin of $|y|$, by counting $\Lambda$ 
[$\bar{\Lambda}$] candidates in the signal region 
(with invariant mass in the range $1.1067$ to $1.1247$ GeV)
and subtracting the corresponding counts in two side band regions
($1.0927$ to $1.1017$ GeV, and $1.1297$ to $1.1387$ GeV).
These four numbers define the 
normalization $N$ and three raw asymmetries $A'_{FB}$, $A'_{NS}$, and $A'_{\Lambda \bar{\Lambda}}$:
\begin{eqnarray}
N_F(\Lambda) & \equiv & N (1 + A'_{FB}) (1 - A'_{NS}) (1 + A'_{\Lambda \bar{\Lambda}}), \nonumber \\
N_B(\Lambda) & \equiv & N (1 - A'_{FB}) (1 + A'_{NS}) (1 + A'_{\Lambda \bar{\Lambda}}), \nonumber \\
N_F(\bar{\Lambda}) & \equiv & N (1 + A'_{FB}) (1 + A'_{NS}) (1 - A'_{\Lambda \bar{\Lambda}}), \nonumber \\
N_B(\bar{\Lambda}) & \equiv & N (1 - A'_{FB}) (1 - A'_{NS}) (1 - A'_{\Lambda \bar{\Lambda}}).
\label{def}
\end{eqnarray}
The asymmetry $A'_{NS}$ measures the relative excess
of reconstructed $\Lambda$'s plus $\bar{\Lambda}$'s with longitudinal momentum in the $\bar{p}$ direction
(north) with respect to the $p$ direction (south).
The asymmetry $A'_{\Lambda \bar{\Lambda}}$ measures the relative excess of 
reconstructed $\Lambda$'s with respect to $\bar{\Lambda}$'s.
The raw asymmetries $A'_{FB}$, $A'_{NS}$, and $A'_{\Lambda \bar{\Lambda}}$ defined in Eq.\ (\ref{def})
have contributions from
the physical processes of the $p \bar{p}$ collisions
($A_{FB}$, $A_{NS}$, and $A_{\Lambda \bar{\Lambda}}$, respectively), and from detector
effects. 
As we discuss below, the raw asymmetries $A'_{NS}$ and $A'_{\Lambda \bar{\Lambda}}$ are dominated by
detector effects, while $A'_{FB}$ is due to the physics of the $p \bar{p}$ 
collisions with negligible contributions from detector effects.
Up to second order terms in the asymmetries, we have,
\begin{eqnarray}
A'_{FB} & = & \frac{N_F(\Lambda) - N_B(\Lambda) + N_F(\bar{\Lambda}) - N_B(\bar{\Lambda})}
{N_F(\Lambda) + N_B(\Lambda) + N_F(\bar{\Lambda}) + N_B(\bar{\Lambda})} \nonumber \\
& & + A'_{NS} A'_{\Lambda \bar{\Lambda}}, \nonumber \\
A'_{NS} & = & \frac{-N_F(\Lambda) + N_B(\Lambda) + N_F(\bar{\Lambda}) - N_B(\bar{\Lambda})}
{N_F(\Lambda) + N_B(\Lambda) + N_F(\bar{\Lambda}) + N_B(\bar{\Lambda})} \nonumber \\
& & + A'_{FB} A'_{\Lambda \bar{\Lambda}}, \nonumber \\
A'_{\Lambda \bar{\Lambda}} & = & \frac{N_F(\Lambda) + N_B(\Lambda) - N_F(\bar{\Lambda}) - N_B(\bar{\Lambda})}
{N_F(\Lambda) + N_B(\Lambda) + N_F(\bar{\Lambda}) + N_B(\bar{\Lambda})} \nonumber \\
& & + A'_{FB} A'_{NS}.
\label{lin}
\end{eqnarray}

The initial $p \bar{p}$ state is invariant with respect to
CP-conjugation. 
Note that CP-conjugation changes the signs of $A_{NS}$ and $A_{\Lambda \bar{\Lambda}}$,
while $A_{FB}$ is left unchanged. A non-zero $A_{NS}$ or $A_{\Lambda \bar{\Lambda}}$ 
would indicate CP-violation.

The raw asymmetry $A'_{NS}$ is different from zero if
the north half of the D0 detector has a different acceptance times
efficiency than the south half of the detector.
This detector asymmetry does not modify $A'_{FB}$ or $A'_{\Lambda \bar{\Lambda}}$
as defined in Eq.\ (\ref{def}).

Antiprotons have a larger inelastic cross section with the
detector material than protons. This difference results in
a higher detection efficiency for $\Lambda$'s than $\bar{\Lambda}$'s.
This difference in efficiencies modifies $A'_{\Lambda \bar{\Lambda}}$ but
does not modify $A'_{FB}$ or $A'_{NS}$
as defined in Eq.\ (\ref{def}).

The solenoid and toroid magnet polarities
are reversed approximately every two weeks during data taking 
so that at each of the
four solenoid-toroid polarity combinations 
approximately the same number of events are collected. 
The raw asymmetries obtained with
each magnet polarity show variations of	up to 
$\pm 0.004$ for $A'_{FB}$, 
$\pm 0.008$ for $A'_{NS}$, and
$\pm 0.003$ for $A'_{\Lambda \bar{\Lambda}}$.
Consider an event with $\Lambda \rightarrow p \pi^-$, and
the charge-conjugate (C) event 
with $\bar{\Lambda} \rightarrow \overline{p} \pi^+$,
with the
same momenta for all corresponding tracks. Assume that, due to some
detector geometric effect, the former event has a larger acceptance times
efficiency than the latter event for a given solenoid and toroid
polarity. Now reverse these polarities. The tracks of the event
$\Lambda \rightarrow p \pi^-$ with one solenoid and toroid
polarity coincide with the tracks of the event
$\bar{\Lambda} \rightarrow \overline{p} \pi^+$ with the opposite
polarities. So with
reversed polarities it is now the event with
$\bar{\Lambda} \rightarrow \overline{p} \pi^+$ that has the
larger acceptance times efficiency.
The conjugation $\Lambda \leftrightarrow \bar{\Lambda}$
reverses the signs of
$A'_{FB}$ and $A'_{\Lambda \bar{\Lambda}}$, and leaves $A'_{NS}$ unchanged.
We conclude that by collecting equal numbers of $\Lambda$ plus
$\bar{\Lambda}$ for each solenoid and toroid magnet polarity
combination,
geometrical detector effects are canceled for 
$A'_{FB}$ and $A'_{\Lambda \bar{\Lambda}}$, but not for $A'_{NS}$ (if C symmetry holds).
We weight events for each polarity combination
to achieve these cancellations. 

We correct $A'_{NS}$ using the
measurements with $K_S$ by setting $A_{NS} = A'_{NS} - A'_{NS}(K_S)$.
None of the detector effects
discussed above affects $A'_{FB}$ as defined in Eq.\ (\ref{def}), 
so we set $A'_{FB} = A_{FB}$ (as a cross-check we verify this equality with $K_S$,
i.e. $A'_{FB}(K_S) = 0$ within statistical uncertainties). 
We  do not measure $A_{\Lambda \bar{\Lambda}}$ as we are not
able to separate the effect due to different
reconstruction efficiencies  from the raw asymmetry $A'_{\Lambda \bar{\Lambda}}$.

\begin{figure}
\begin{center}
\scalebox{0.35}
{\includegraphics{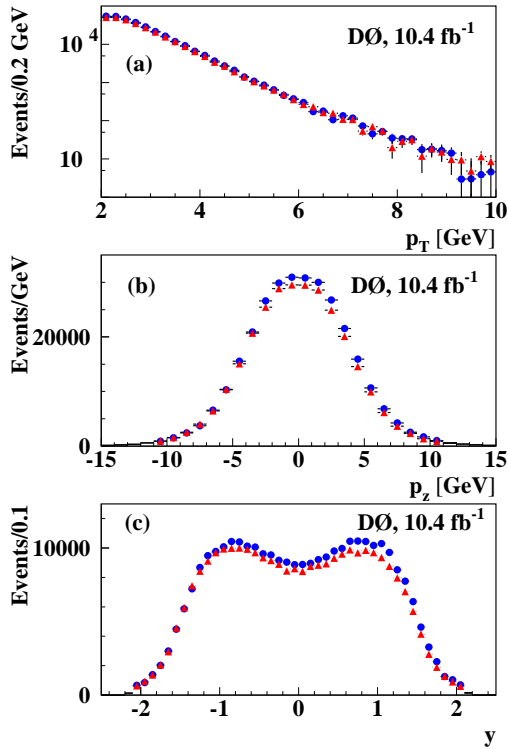}}
\caption{
(color online)
Distributions of (a) $p_T$, (b) $p_z$, and (c) $y$
of reconstructed $\Lambda$'s (blue circles) and $\bar{\Lambda}$'s (red triangles)
with $p_T > 2.0$ GeV, for the minimum bias data sample
$p \bar{p} \rightarrow \Lambda (\bar{\Lambda}) X$.
}
\label{pT_pz_eta_3}
\end{center}
\end{figure}

\begin{figure}
\begin{center}
\scalebox{0.35}
{\includegraphics{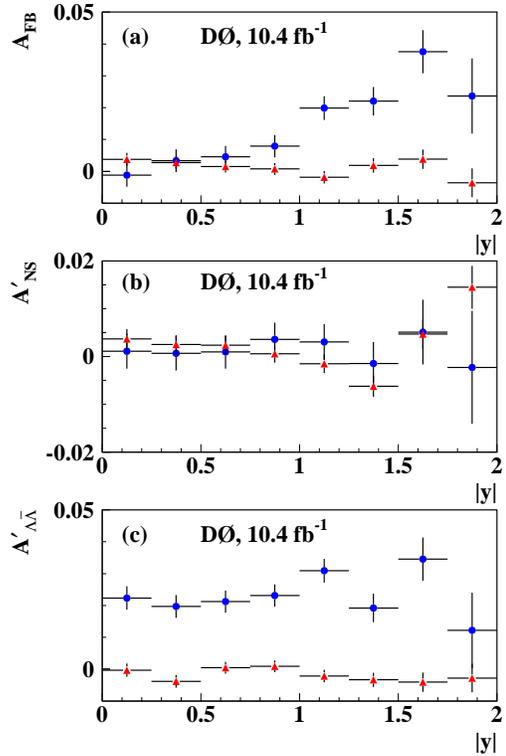}}
\caption{
(color online)
Asymmetries (a) $A_{FB} = A'_{FB}$, (b) $A'_{NS}$, and (c) $A'_{\Lambda \bar{\Lambda}}$ 
of reconstructed $\Lambda$ and $\bar{\Lambda}$
(blue circles) and $K_S$ (red triangles)
with $p_T > 2.0$ GeV, as functions of $|y|$,
for the minimum bias data samples
$p \bar{p} \rightarrow \Lambda (\bar{\Lambda}) X$
or $p \bar{p} \rightarrow K_S X$
respectively.
Uncertainties are statistical.
}
\label{A_AFB_ANS_3}
\end{center}
\end{figure}

\section{Results for minimum bias events}

\begin{figure}
\begin{center}
\scalebox{0.35}
{\includegraphics{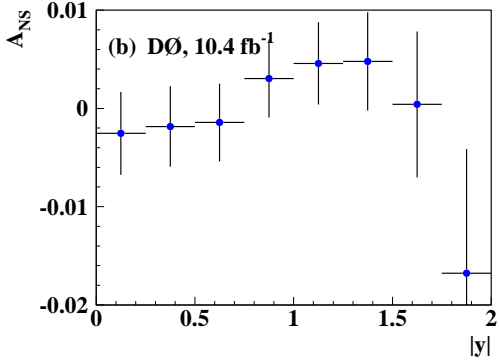}}
\caption{
Corrected asymmetry $A_{NS} = A'_{NS} - A'_{NS}(K_S)$ of $\Lambda$ and $\bar{\Lambda}$
with $p_T > 2.0$ GeV, as a function of $|y|$,
for the minimum bias data sample
$p \bar{p} \rightarrow \Lambda (\bar{\Lambda}) X$.
Uncertainties are statistical.
}
\label{AFB_ANS_4}
\end{center}
\end{figure}

We now consider minimum bias events 
$p \bar{p} \rightarrow \Lambda (\bar{\Lambda}) X$ and
the control sample $p \bar{p} \rightarrow K_S X$.
Distributions of $p_T$, $p_z$, and $y$ of reconstructed $\Lambda$'s and
$\bar{\Lambda}$'s are shown in Fig.\ \ref{pT_pz_eta_3}.
The raw asymmetries of $\Lambda$, $\bar{\Lambda}$ and $K_S$
for $p_T > 2.0$ GeV are presented in Fig.\ \ref{A_AFB_ANS_3}.
We expect the asymmetries $A'_{FB}(K_S)$ and $A'_{\Lambda \bar{\Lambda}}(K_S)$ to
be zero, while $A'_{NS}(K_S)$ is not necessarily zero.
These expectations are satisfied within the statistical
uncertainties.
From Fig.\ \ref{A_AFB_ANS_3} (c) we obtain $A'_{\Lambda \bar{\Lambda}} \approx 0.022$.
This asymmetry is different from zero
as expected from the different
inelastic cross-sections of $p$ and $\bar{p}$, and of 
$\Lambda$ and $\bar{\Lambda}$, with the detector material.
The asymmetries in Fig.\ \ref{A_AFB_ANS_3} were 
obtained from Eqs.\ (\ref{lin}) but neglecting
the quadratic terms. Therefore the forward-backward
asymmetries shown in Fig.\ \ref{A_AFB_ANS_3} need
corrections $A'_{NS} A'_{\Lambda \bar{\Lambda}}$ due to detector effects.
These corrections, obtained  bin-by-bin from Fig.\ \ref{A_AFB_ANS_3} (b) and (c), are measured
to be consistent with zero within their statistical uncertainties.
As they are small, they are not applied as corrections, but are
treated as systematic uncertainties. They vary
from $\pm 0.0001$ for the first bin of $|y|$
to $\pm 0.0004$ for the $1.5 < |y| < 1.75$ bin.
The results for $A_{FB}$ are presented in Fig.\ \ref{A_AFB_ANS_3} and
Table \ref{afb_zb_mu}. 
The corrected 
asymmetry $A_{NS} = A'_{NS} - A'_{NS}(K_S)$ is consistent with 
zero within the statistical uncertainties, 
so we observe no significant CP violation in $A_{NS}$,
as shown in Fig.\ \ref{AFB_ANS_4}.

\begin{figure}
\begin{center}
\scalebox{0.35}
{\includegraphics{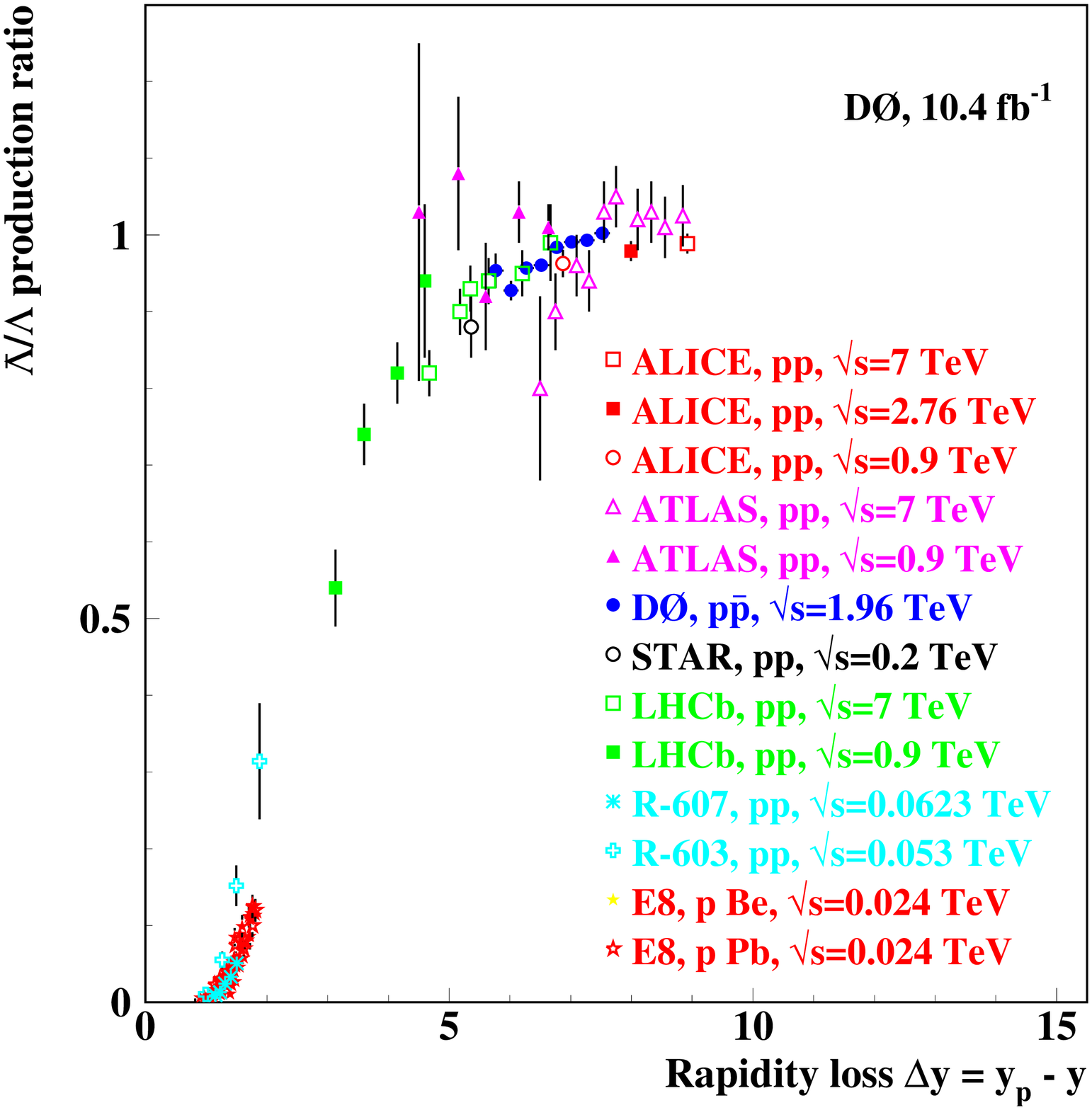}}
\caption{$\bar{\Lambda}/\Lambda$ production ratio
as a function of the rapidity loss $\Delta y \equiv y_p - y$
for several experiments that study
reactions $p Z \rightarrow \Lambda (\bar{\Lambda}) X$
for targets $Z = p, \bar{p}$, Be and Pb.
The experiments are
ALICE \cite{ALICE},
ATLAS \cite{ATLAS},
D0  (this analysis),
STAR \cite{STAR},
LHCb \cite{LHCb}, 
ISR R-607 \cite{ISR607}, 
ISR R-603 \cite{ISR}, and
the fixed target experiment
Fermilab E8 studying $p$-\textrm{Be} and $p$-\textrm{Pb} collisions at a
beam energy of 300 GeV \cite{E8}.
}
\label{comparison}
\end{center}
\end{figure}

\begin{figure}
\begin{center}
\scalebox{0.35}
{\includegraphics{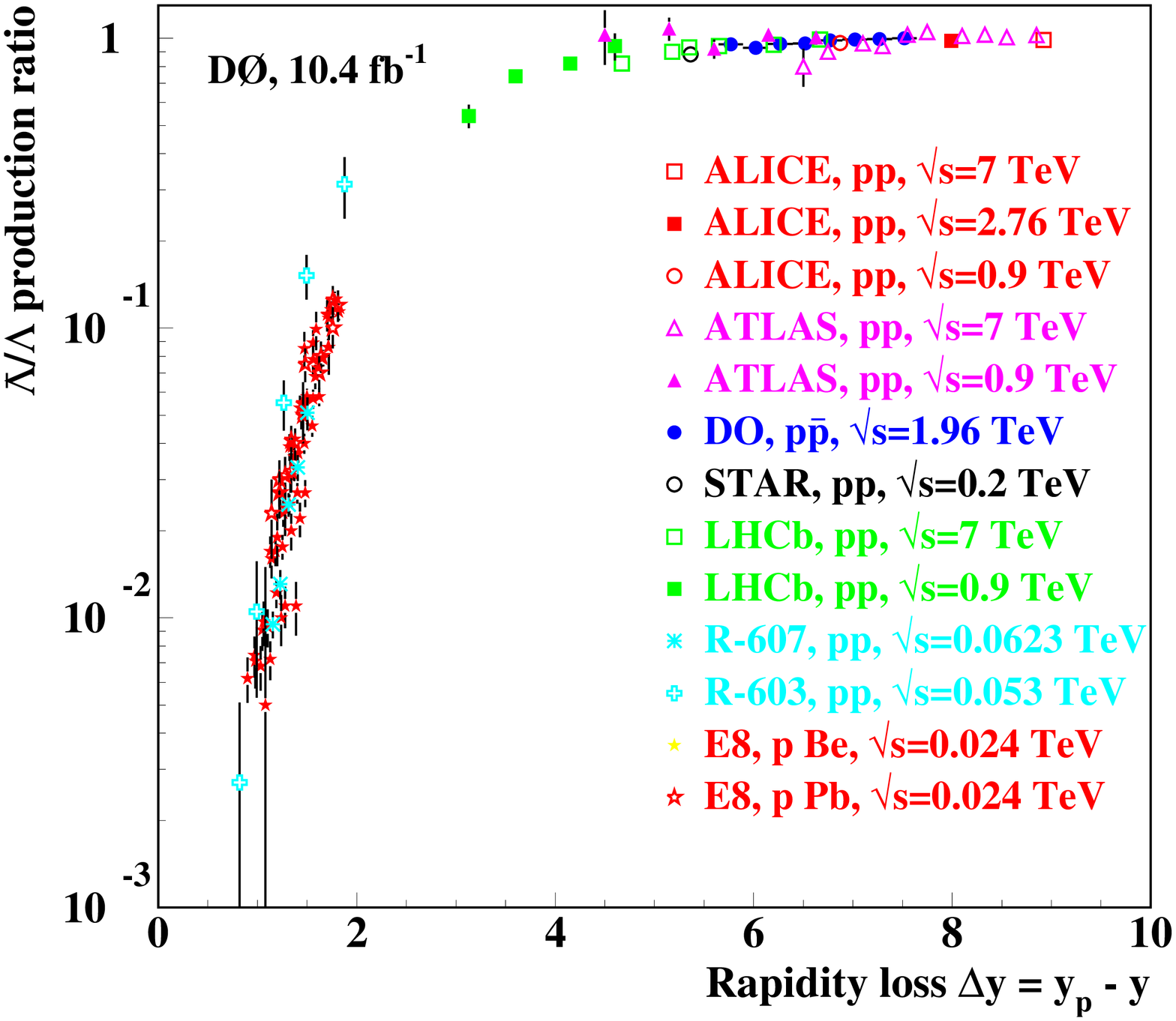}}
\caption{
Same as Fig.\ \ref{comparison} with logarithmic scale.
}
\label{comparison_log}
\end{center}
\end{figure}

In Figs.\ \ref{comparison} and \ref{comparison_log},
the asymmetry $A_{FB}$ shown in Fig.\ \ref{A_AFB_ANS_3}
is compared
with other experiments that study collisions
$p Z \rightarrow \Lambda (\bar{\Lambda}) X$ for several targets $Z = p, \bar{p}$, Be and Pb.
For the D0 minimum bias data in Figs.\ \ref{comparison}
and \ref{comparison_log}, we plot 
$[\sigma_B(\Lambda) + \sigma_B(\bar{\Lambda})]/[\sigma_F(\Lambda) + \sigma_F(\bar{\Lambda})]
= (1 - A_{FB})/(1 + A_{FB})$.
We should note that the point $y = 0$ in the center of mass
for $p \bar{p}$ collisions has a $\bar{\Lambda}/\Lambda$ production ratio
equal to 1 if CP is conserved, which is not necessarily the case for $p p$ collisions,
so this D0 point at large rapidity loss should be excluded from
the comparison with $p p$ data.
From Figs.\ \ref{comparison} and \ref{comparison_log} we conclude that the
$\bar{\Lambda}/\Lambda$ production ratio 
is approximately
a universal function of the ``rapidity loss"
$\Delta y \equiv y_p - y$, independent of $\sqrt{s}$ or target $Z$.
Here $y_p$ is the rapidity of the proton beam,
and $y$ is the rapidity	of the $\Lambda$ or $\bar{\Lambda}$.


\begin{table*}
\caption{\label{afb_zb_mu}
Forward-backward asymmetry $A_{FB}$ of
$\Lambda$ and $\bar{\Lambda}$ with $p_T > 2.0$ GeV
in minimum bias events
$p \bar{p} \rightarrow \Lambda (\bar{\Lambda}) X$,
events $p \bar{p} \rightarrow J/\psi \Lambda (\bar{\Lambda}) X$, 
and events $p \bar{p} \rightarrow \mu^\pm \Lambda (\bar{\Lambda}) X$.
The first uncertainty is statistical, the second is
systematic.
}
\begin{ruledtabular}
\newcolumntype{A}{D{A}{\pm}{-0.5}}
\begin{tabular}{cAAc}
$|y|$ & \multicolumn{1}{c}{$A_{FB} \times 100$ (min. bias)} & \multicolumn{1}{c}{$A_{FB} \times 100$ (with $J/\psi$)}
  & $A_{FB} \times 100$ (with $\mu$) \\
\hline
0.00 to 0.25 & -0.12\ A \ 0.37 \pm 0.01 & -0.21\ A \ 0.58 \pm 0.01 
  & $0.16 \pm 0.09 \pm 0.02$ \\
0.25 to 0.50 & 0.33\ A \ 0.36 \pm 0.01  & 0.10\ A \ 0.57 \pm 0.02
  & $0.24 \pm 0.09 \pm 0.02$ \\
0.50 to 0.75 & 0.45\ A \ 0.35 \pm 0.01  & 0.69\ A \ 0.56 \pm 0.02
  & $0.67 \pm 0.08 \pm 0.02$ \\
0.75 to 1.00 & 0.79\ A \ 0.35 \pm 0.02  & 0.55\ A \ 0.56 \pm 0.02
  & $0.85 \pm 0.08 \pm 0.02$ \\
1.00 to 1.25 & 1.99\ A \ 0.37 \pm 0.02  & 0.69\ A \ 0.59 \pm 0.03
  & $1.57 \pm 0.09 \pm 0.02$ \\
1.25 to 1.50 & 2.20\ A \ 0.45 \pm 0.02  & 1.72\ A \ 0.72 \pm 0.03
  & $1.98 \pm 0.10 \pm 0.04$ \\
1.50 to 1.75 & 3.75\ A \ 0.68 \pm 0.03  & 3.24\ A \ 1.12 \pm 0.06
  & $2.53 \pm 0.16 \pm 0.06$ \\
1.75 to 2.00 & 2.37\ A \ 1.18 \pm 0.04  & 2.64\ A \ 2.06 \pm 0.06
  & $3.11 \pm 0.30 \pm 0.06$ \\
\end{tabular}
\end{ruledtabular}
\end{table*}

\begin{table*}
\caption{\label{afb_mu}
Forward-backward asymmetry $A_{FB}$ of
$\Lambda$ and $\bar{\Lambda}$ in bins of $p_T$
in events
$p \bar{p} \rightarrow \mu^\pm \Lambda (\bar{\Lambda}) X$.
The first uncertainty is statistical, the second is
systematic.
}
\begin{ruledtabular}
\newcolumntype{A}{D{A}{\pm}{-0.5}}
\begin{tabular}{cAAA}
$|y|$ & \multicolumn{1}{c}{$A_{FB} \times 100$} & \multicolumn{1}{c}{$A_{FB} \times 100$} 
  & \multicolumn{1}{c}{$A_{FB} \times 100$}  \\
 & \multicolumn{1}{c}{$2 < p_T < 4$ GeV} & \multicolumn{1}{c}{$4 < p_T < 6$ GeV} 
  & \multicolumn{1}{c}{$p_T > 6$ GeV} \\
\hline
0.00 to 0.25 & 0.21\ A \ 0.09 \pm 0.02 & -0.27\ A \ 0.28 \pm 0.02
  & 0.57\ A \ 0.69 \pm 0.02 \\
0.25 to 0.50 & 0.25\ A \ 0.09 \pm 0.02 & 0.20\ A \ 0.27 \pm 0.02
  & -0.47\ A \ 0.63 \pm 0.02 \\
0.50 to 0.75 & 0.70\ A \ 0.08 \pm 0.02 & 0.50\ A \ 0.26 \pm 0.02
  & 1.11\ A \ 0.58 \pm 0.02 \\
0.75 to 1.00 & 0.82\ A \ 0.08 \pm 0.02 & 1.02\ A \ 0.25 \pm 0.02
  & 0.57\ A \ 0.54 \pm 0.02 \\
1.00 to 1.25 & 1.60\ A \ 0.10 \pm 0.02 & 1.39\ A \ 0.25 \pm 0.02
  & 2.38\ A \ 0.52 \pm 0.02 \\
1.25 to 1.50 & 1.94\ A \ 0.11 \pm 0.04 & 2.17\ A \ 0.27 \pm 0.04
  & 2.43\ A \ 0.57 \pm 0.04 \\
1.50 to 1.75 & 2.61\ A \ 0.17 \pm 0.06 & 2.10\ A \ 0.42 \pm 0.06
  & 4.77\ A \ 0.85 \pm 0.06 \\
1.75 to 2.00 & 3.05\ A \ 0.32 \pm 0.06 & 3.49\ A \ 0.83 \pm 0.06
  & 6.32\ A \ 1.69 \pm 0.06 \\
\end{tabular}
\end{ruledtabular}
\end{table*}

\section{Results for events with a $J/\psi$ or a muon}

\begin{figure}
\begin{center}
\scalebox{0.35}
{\includegraphics{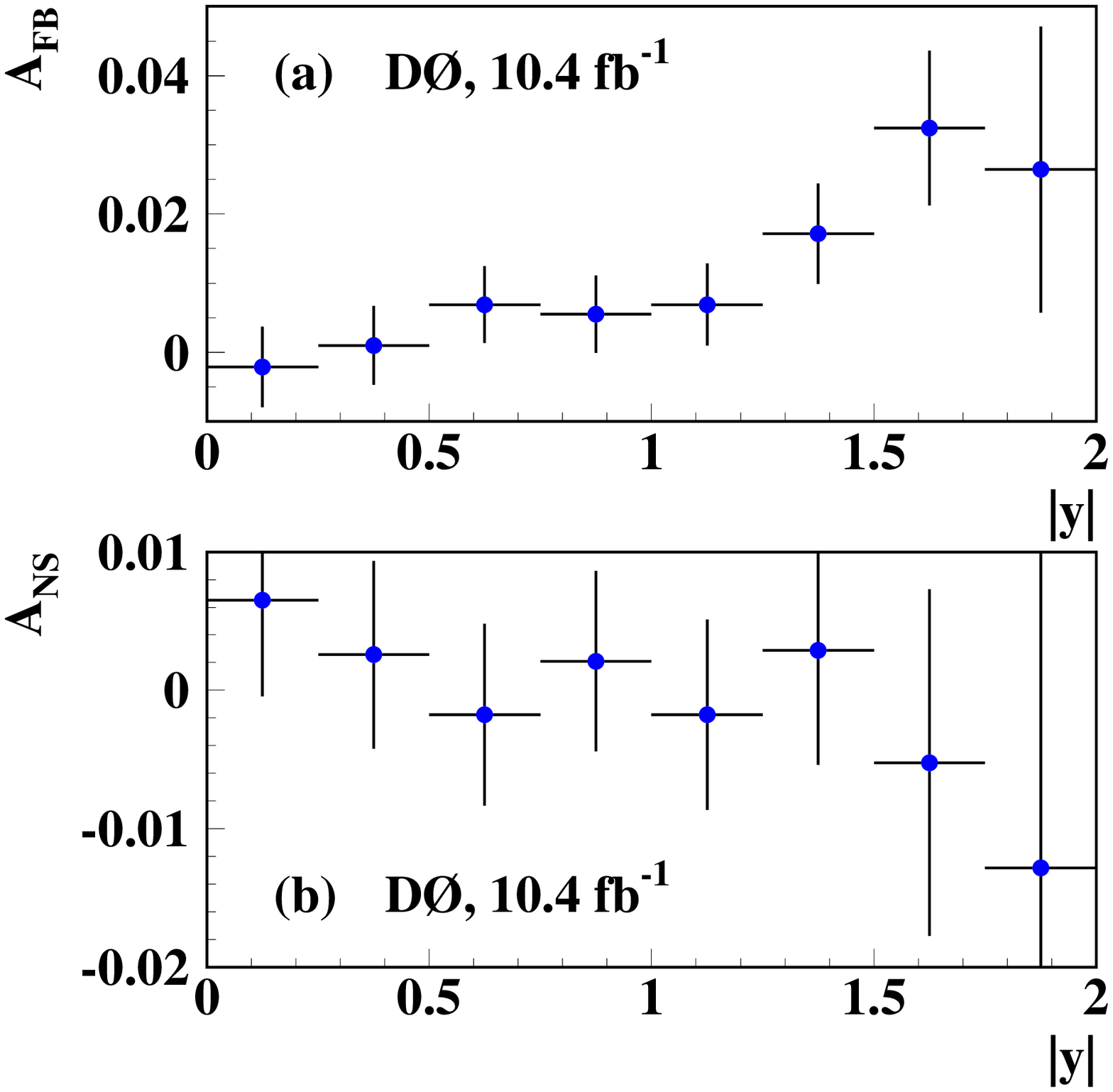}}
\caption{
Asymmetries (a) $A_{FB} = A'_{FB}$ and (b) $A_{NS} = A'_{NS} - A'_{NS}(K_S)$ 
of $\Lambda$ and $\bar{\Lambda}$
with $p_T > 2.0$ GeV, as functions of $|y|$,
for the data sample
$p \bar{p} \rightarrow J/\psi \Lambda (\bar{\Lambda}) X$.
Uncertainties are statistical.
}
\label{AFB_ANS_7}
\end{center}
\end{figure}

The results of the measurements with the data set
$p \bar{p} \rightarrow J/\psi \Lambda (\bar{\Lambda}) X$
are presented in Fig.\ \ref{AFB_ANS_7} and Table \ref{afb_zb_mu}.
We note that $A_{NS}$ is consistent with zero, whereas $A_{FB}$ is
significantly nonzero at large $|y|$.

\begin{figure}
\begin{center}
\scalebox{0.35}
{\includegraphics{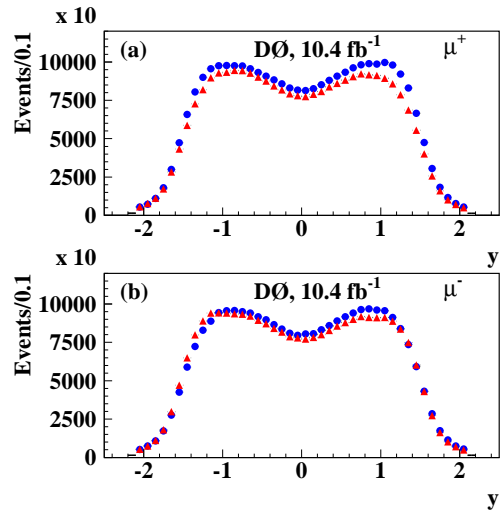}}
\caption{
(color online)
Distributions of rapidity $y$ of reconstructed 
$\Lambda$'s (blue circles) and $\bar{\Lambda}$'s
(red triangles) for events with (a) $\mu^+$ or (b)
$\mu^-$, for $p_T > 2.0$ GeV, for events
$p \bar{p} \rightarrow \mu^\pm \Lambda (\bar{\Lambda}) X$.
}
\label{y_l_lbar_mup_mum_2_114_2}
\end{center}
\end{figure}

\begin{figure}
\begin{center}
\scalebox{0.35}
{\includegraphics{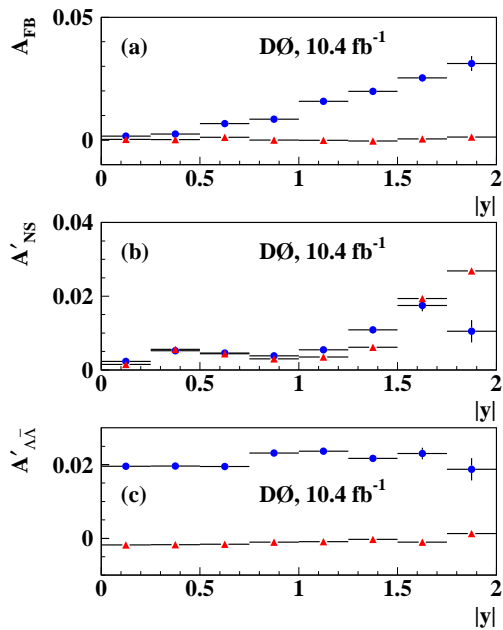}}
\caption{
(color online)
Asymmetries (a) $A_{FB} = A'_{FB}$, (b) $A'_{NS}$, and (c) $A'_{\Lambda \bar{\Lambda}}$ 
of reconstructed $\Lambda$ and $\bar{\Lambda}$
(blue circles) and $K_S$ (red triangles)
with $p_T > 2.0$ GeV, as functions of $|y|$,
for events
$p \bar{p} \rightarrow \mu^\pm \Lambda (\bar{\Lambda}) X$
or $p \bar{p} \rightarrow \mu^\pm K_S X$ respectively.
Uncertainties are statistical.
}
\label{A_AFB_ANS_mupm}
\end{center}
\end{figure}

\begin{figure}
\begin{center}
\scalebox{0.35}
{\includegraphics{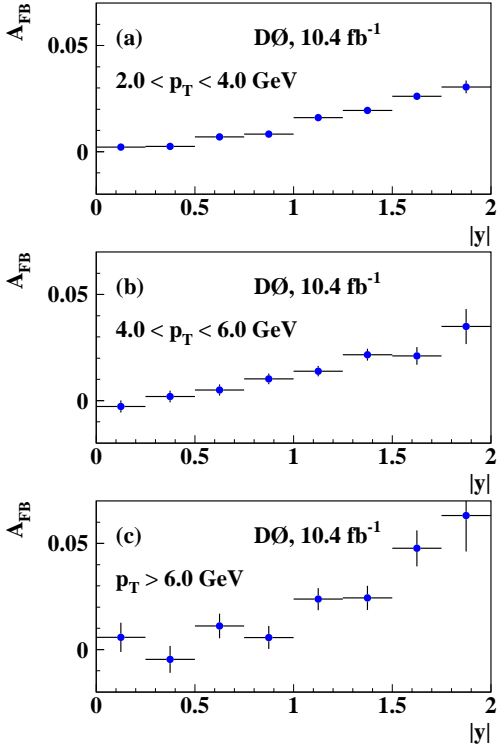}}
\caption{
Asymmetry $A_{FB}$ as a function of $|y|$
for events $p \bar{p} \rightarrow \mu^\pm \Lambda (\bar{\Lambda}) X$
for (a) $2.0 < p_T < 4.0$ GeV, (b) $4.0 < p_T < 6.0$ GeV, and (c) $p_T > 6.0$ GeV.
Uncertainties are statistical. 
}
\label{afb_2_4_6_mup_mum_6_114_2}
\end{center}
\end{figure}

We now consider the large data sample
$p \bar{p} \rightarrow \mu^\pm \Lambda (\bar{\Lambda}) X$.
Rapidity distributions for reconstructed $\Lambda$'s and $\bar{\Lambda}$'s
are presented in Fig.\ \ref{y_l_lbar_mup_mum_2_114_2}.
After accounting for the different efficiencies to detect
$\Lambda$ and $\bar{\Lambda}$,
we find that there are more
events $\Lambda \mu^+$ and $\bar{\Lambda} \mu^-$, than 
events $\Lambda \mu^-$ and $\bar{\Lambda} \mu^+$.
Examples of decays with a $\Lambda \mu^+$ correlation are:
$\Lambda_c^+ \rightarrow \Lambda \mu^+ \nu_\mu$ and
$p \bar{p} \rightarrow \Lambda K^+ X$  
followed by $K^+ \rightarrow \mu^+ \nu_\mu$
(note that the $\Lambda$ and $K^+$ share an $s \bar{s}$ pair).
The reverse $\Lambda \mu^-$ correlation occurs for
$\Lambda_b \rightarrow \mu^- \Lambda_c^+ \bar{\nu}_\mu X$ 
with $\Lambda_c^+ \rightarrow \Lambda X$.
Measurements of $A_{FB}(|y|)$ for events with $\mu^+$ or $\mu^-$
are found to be consistent within statistical uncertainties, so we combine
events with $\mu^+$ and $\mu^-$
and obtain the results presented in Fig.\ \ref{A_AFB_ANS_mupm}.
We assign to $A_{FB}$ a systematic uncertainty equal to
the entire detector effect, $A'_{NS} A'_{\Lambda \bar{\Lambda}}$. 
Numerical results
are presented in Table \ref{afb_zb_mu}.
The forward-backward asymmetry $A_{FB}$ as a 
function of $|y|$ for different lambda transverse momentum
bins is shown in Fig.\ \ref{afb_2_4_6_mup_mum_6_114_2}
and Table \ref{afb_mu}.
Note that $A_{FB}$ is only weakly dependent on $p_T(\Lambda)$.

\begin{figure}
\begin{center}
\scalebox{0.35}
{\includegraphics{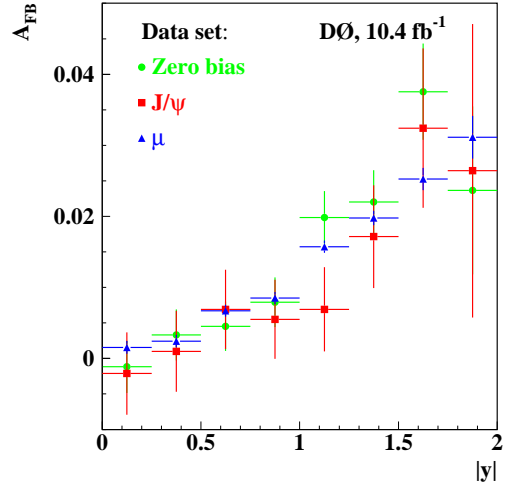}}
\caption{
(color online)
Asymmetry $A_{FB}$ as a function of $|y|$
for events 
$p \bar{p} \rightarrow \Lambda (\bar{\Lambda}) X$ (green circles),
$p \bar{p} \rightarrow J/\psi \Lambda (\bar{\Lambda}) X$ (red squares), and
$p \bar{p} \rightarrow \mu^\pm \Lambda (\bar{\Lambda}) X$ (blue triangles)
for $p_T > 2.0$ GeV. Uncertainties are statistical.
}
\label{A_FB_three}
\end{center}
\end{figure}

The final results of this analysis are summarized in
Tables \ref{afb_zb_mu} and \ref{afb_mu}, and Figs. 
\ref{afb_2_4_6_mup_mum_6_114_2} and \ref{A_FB_three}.

\section{Conclusions}
We have measured the forward-backward asymmetry of $\Lambda$ and $\bar{\Lambda}$
production $A_{FB}$ as a function of rapidity $|y|$
for three data sets:
$p \bar{p} \rightarrow \Lambda (\bar{\Lambda}) X$,
$p \bar{p} \rightarrow J/\psi \Lambda (\bar{\Lambda}) X$, and
$p \bar{p} \rightarrow \mu^\pm \Lambda (\bar{\Lambda}) X$.
The asymmetry $A_{FB}$ is a function of $|y|$ that
does not depend significantly on
the data set or data composition (see Fig.\ \ref{A_FB_three}),
and is weakly dependent on $p_T$ (see Fig.\ \ref{afb_2_4_6_mup_mum_6_114_2}).
The measurement of $A_{FB}$
in $p \bar{p}$ collisions can be compared with
the $\bar{\Lambda}/\Lambda$ production ratio 
measured by a wide range of proton scattering experiments.
This production ratio is confirmed to be approximately a universal function of the
``rapidity loss" $y_p - y$,
that does not depend significantly (or depends only weakly) on
the total center of mass energy $\sqrt{s}$ or
target (see Figs.\ \ref{comparison} and \ref{comparison_log}).
This result supports the view that a strange quark produced directly in the
hard scattering of point-like partons,
or indirectly in the subsequent showering, 
can coalesce with a diquark remnant
of the beam particle to produce a lambda with a probability
that increases as the rapidity difference between the proton
and lambda decreases.

\begin{acknowledgments}
%

We thank the staffs at Fermilab and collaborating institutions,
and acknowledge support from the
Department of Energy and National Science Foundation (United States of America);
Alternative Energies and Atomic Energy Commission and
National Center for Scientific Research/National Institute of Nuclear and Particle Physics  (France);
Ministry of Education and Science of the Russian Federation, 
National Research Center ``Kurchatov Institute" of the Russian Federation, and 
Russian Foundation for Basic Research  (Russia);
National Council for the Development of Science and Technology and
Carlos Chagas Filho Foundation for the Support of Research in the State of Rio de Janeiro (Brazil);
Department of Atomic Energy and Department of Science and Technology (India);
Administrative Department of Science, Technology and Innovation (Colombia);
National Council of Science and Technology (Mexico);
National Research Foundation of Korea (Korea);
Foundation for Fundamental Research on Matter (The Netherlands);
Science and Technology Facilities Council and The Royal Society (United Kingdom);
Ministry of Education, Youth and Sports (Czech Republic);
Bundesministerium f\"{u}r Bildung und Forschung (Federal Ministry of Education and Research) and 
Deutsche Forschungsgemeinschaft (German Research Foundation) (Germany);
Science Foundation Ireland (Ireland);
Swedish Research Council (Sweden);
China Academy of Sciences and National Natural Science Foundation of China (China);
and
Ministry of Education and Science of Ukraine (Ukraine).
%
\end{acknowledgments}

\end{document}